\documentclass[prl,showclass, twocolumn,preprintnumbers,amsmath,amssymb]{revtex4-2}
\usepackage{graphicx}% Include figure files
\usepackage{dcolumn}% Align table columns on decimal point

\begin{document}

\title{Ising models of deep neural networks}

\author{Dusan Stosic}
\email{dbstosic@gmail.com}
\author{Darko Stosic}
\email{ddstosic@bu.edu}
\affiliation{Centro de Inform\'atica, Universidade Federal de Pernambuco, Av. Luiz Freire s/n, 50670-901, Recife, PE, Brazil}
\author{Borko Stosic}
\email{borko.stosic@ufrpe.br}
\affiliation{
Departamento de Estat\' \i stica e Inform\' atica, 
Universidade Federal Rural de Pernambuco,\\
Rua Dom Manoel de Medeiros s/n, Dois Irm\~ aos,
52171-900 Recife-PE, Brazil
}

\date{\today}% It is always \today, today,

\begin{abstract}
This work maps deep neural networks to classical Ising spin models, allowing them to be described using statistical thermodynamics. The density of states shows that structures emerge in the weights after they have been trained -- well-trained networks span a much wider range of realizable energies compared to poorly trained ones. These structures propagate throughout the entire network and are not observed in individual layers.  The energy values correlate to performance on tasks, making it possible to distinguish networks based on quality without access to data. Thermodynamic properties such as specific heat are also studied, revealing a higher critical temperature in trained networks.
\end{abstract}


\maketitle

\section{Introduction}
Deep learning has emerged as a disruptive technology capable of solving complex problems in vision~\cite{krizhevsky2012,gaugan,dalle}, language~\cite{brown2020}, and even science~\cite{reichstein2019,alphafold,choudhary2022}. Dramatic advances brought by model size~\cite{kaplan2020,henighan2020} spurred an arms race towards training larger deep neural networks, that have grown to contain hundreds of billions of weights~\cite{hestness2017,brown2020,fedus2021,megatron530b}, consuming vasts amount of resources to train and run tasks~\cite{thompson2020}. Thus far, the primary interest in deep learning has been on practical and industrial applications, such as compression techniques~\cite{sparsity,fp16,fp8} to reduce costs, driven by experimental and applied sciences. This is akin to the early days of steam engines, where industrial necessity set the stage for a more robust theory of thermodynamics that was developed over a century later~\cite{hunt2010}. While foundational research for deep neural networks exists~\cite{spectral,bntheory,statmechdl,poggio2020}, the underlying mechanisms that makes them perform well on specific tasks continue to elude us. For example, there remains no clear way to distinguish a well- from a poorly-trained network other than through evaluation on tasks, which by itself can be insufficient to determine whether a network has been properly trained.

Deep neural networks are by construction reminiscent of magnetic model systems where nodes are connected by couplings (weights), giving rise to collective behavior that cannot be described by their individual parts. Similar analogies drawn in other areas of science have motivated the search for ``universal" properties emerging in complex systems found in finance~\cite{plerou2001}, geology~\cite{geology}, social networks~\cite{barabasi1999}, and many more. 
Thus, the powerful formalism of statistical mechanics may be employed to study neural networks in the quest of understanding their mechanics from unique properties that define them, even if large deep neural networks capable of solving complex problems are composed of hundreds of billions of weights, while classical model systems are typically studied on regular grids, with limited connectivity. 

A recent line of research attempts the above approach by  borrowing ideas used to study complex systems. Refs.~\cite{rmtnature,rmtjmlr} employ concepts from random matrix theory, which originates from nuclear physics~\cite{Wigner}, to identify when a neural network might have problems during training. Refs.~\cite{gabrie2018,goldfeld2019} examine neural networks using information-theoretic quantities, like entropy and mutual information. A shortcoming of these works is that they do not take into account structure of the network architecture, but rather analyze each layer individually, thereby potentially missing on critical information about its topology.

This work borrows the concepts of statistical physics (and thermodynamics) to analyze deep neural networks by mapping them to a well-known problem in statistical physics, the Ising model. With this formulation, the weights of a neural network are taken to represent exchange interactions between spins represented by the nodes of the network, and the system can be studied using various properties of spin glass models. The density of states proves particularly effective at uncovering the presence of structures in neural networks after training. These structures are present across a suite of pretrained transformers and shown to correlate with performance on tasks, which suggests an emerging behavior after training that is characteristic of complex systems.

\section{From Deep Neural Networks to Ising Models}
In its simplest form, a deep neural network consists of a set of $L$ layers, where each layer $\ell$ contains $m$ nodes that are connected to $n$ nodes in its neighboring layer $\ell+1$ through weights $w_\ell$ expressed as a $m\times n$ matrix. The goal of training is to learn values of the weights that minimize some energy function $f(x;w_1,\dots,w_L)$ to capture correlations in the data $x$. After training, weights can perform tasks by mapping inputs to desired outputs (e.g., images into labels, questions into answers, translation between languages, etc).

\begin{figure}[htb]
\begin{tabular}{cc}
\includegraphics[trim=100 100 400 50, clip, width=\linewidth]{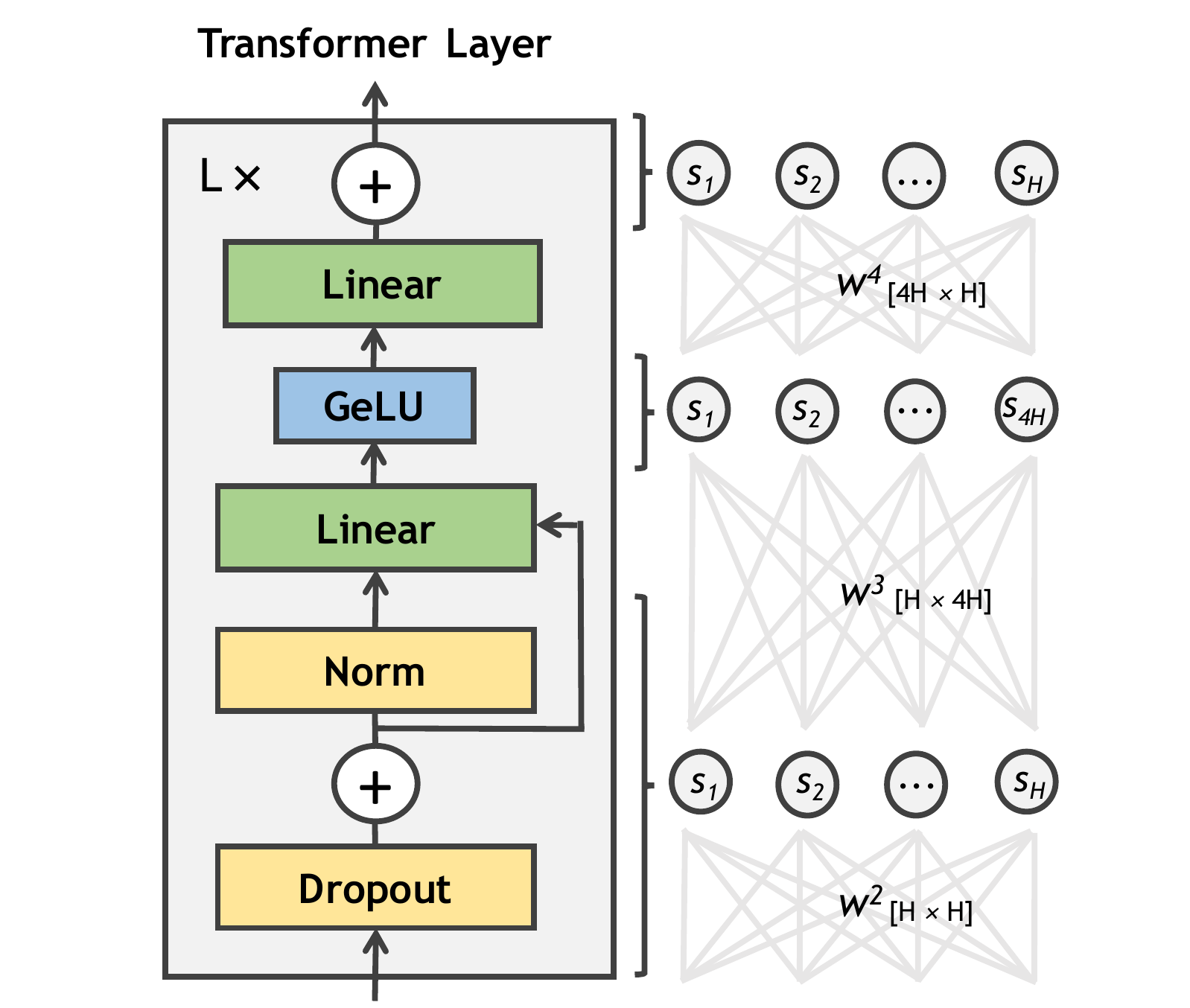}
\end{tabular}
\caption{Illustration of how a transformer layer is mapped into an Ising model spin glass, where weights (e.g, Linear layers) denote exchange couplings, and spins represent neurons spanning multiple activation layers (e.g., Dropout, Add, and Norm).}
\label{figising}
\end{figure}

\begin{figure*}[t]
\begin{tabular}{c}
\includegraphics[width=\linewidth]{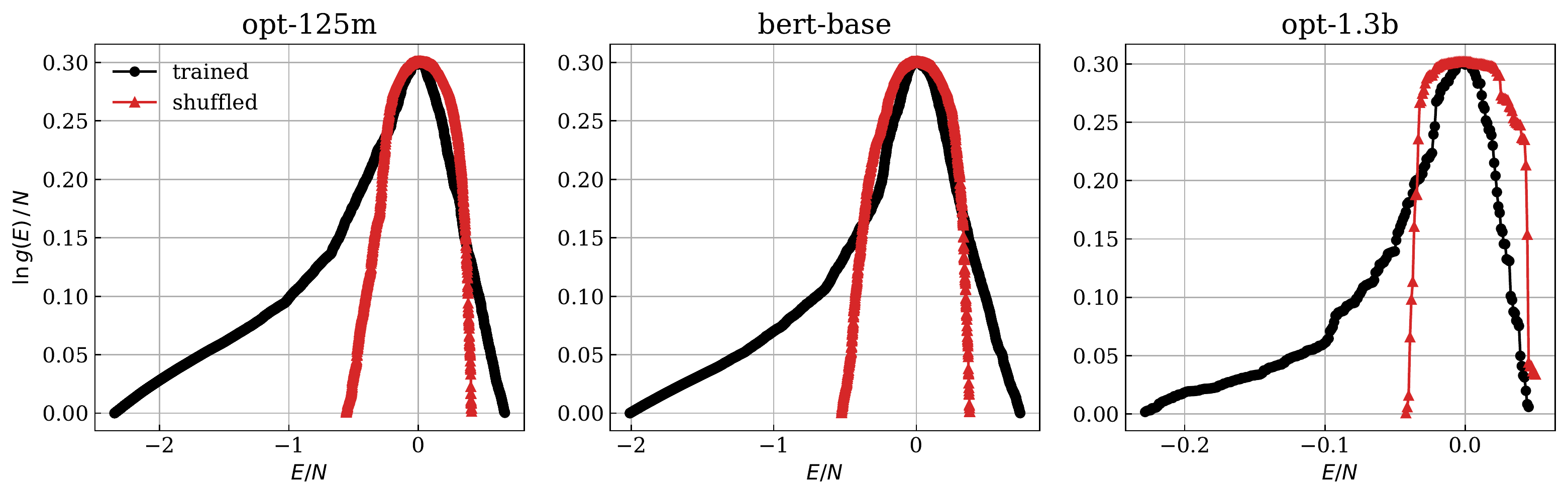}
\end{tabular}
\caption{The density of states, $g(E)$, for a range of transformer networks (of different sizes) pretrained on different data and tasks, and after their values have been shuffled, representing untrained networks without changing the distribution of weights.}
\label{figdos}
\end{figure*}

We propose representing neural networks as Ising models, where each neuron is viewed as a spin with two possible states, an ``up'' or a ``down'' orientation, and each spin pair between consecutive layers interacts via an exchange coupling $J$ that corresponds to the weight matrices $w$. For instance, given a neural network with $y=w_\ell x$, the neuron layers $x$ and $y$ form $m$ and $n$ spins respectively, where spins $x_i$ and $y_j  $ are connected through a coupling $J_{ij}\equiv w_{ij}$. Moreover, all neuron layers between a pair of weights are treated as a single set of spins, e.g. given a neural network with $y=w_2g(f(w_1x))$, the in-between layers $f(\cdot)$ and $g(\cdot)$ are collapsed, forming three unique sets of neurons: $x$, $y$, and $z=g(f(\cdot))$. Since weights can take any positive or negative real value, the resulting ``magnetic system'' can be interpreted as a spin glass with quenched and disordered spin interactions. However, different from most spin models addressed in the literature, neural networks have a large degree, where each node connects to $n$ nearest neighbors instead of typically double the dimension in regular lattices (e.g. $4$ for 2d square, or $6$ for 3d cubic lattice nearest neighbor models).

For this work, neural networks based on transformer architectures~\cite{vaswani2017} are considered. Linear weights in the transformer layers constitute exchange interactions, and activations between those layers are collapsed together into spins, as Figure~\ref{figising} illustrates. Weights for query, key, and value projections are summed, since they eventually merge into the same spins, while other network weights such as biases and normalizations are ignored, including residual connections. Thus, for a neural network of $L$ transformer layers, the spin system consists of $4L+1$ sets of spins, where the number of spins in each set is a multiple of the network size, $n_\ell\propto H$, which are connected to spins in the nearest neighboring layer, amounting to a total of $N=H(7L+1)$ spins and $B=10H^2L$ bonds.

Using this formulation, neural networks can be defined through the Hamiltonian:
\begin{align}
E &= -\sum_{<i,j>}J_{ij}S_iS_j \nonumber \\
&=-\sum_{\ell=1}^{4L}\sum_{i=1}^{n_\ell}\sum_{j=1}^{n_{\ell+1}}J_{ij}^{\ell}S_i^{\ell}S_j^{\ell+1},
\end{align}
where $<>$ denotes summation pairs in neighboring layers, $J\equiv w$ corresponds to the exchange coupling (or weights), and $S_i=\pm1$ is the spin (or neurons) at site $i$. In other words, the Hamiltonian is a summation over neural network weights, where each weight makes a positive or negative contribution to the sum based on values of the neurons it connects.

\section{Calculating the Density of States}
After mapping a neural network to a spin glass model, a number of thermodynamic variables can be used to study it. Many such variables are determined from suitable derivatives of the partition function, which in turn is computed from the density of states. The density of states of a system describes the number $g(E)$ of states (spin configurations) that are accessible to the system at a particular energy level $E$. Since the density of state curves depend on the topology of the lattice alone, they make an ideal candidate for analyzing structures in neural networks.

Wang-Landau algorithm~\cite{wanglandau} has proved to be the most successful approach for estimating the density of states. The algorithm conducts an iterative procedure via a random walk which produces a flat histogram in energy space, and is succintly described as follows. First, the density of states is initialized to $g(E)=1$ for all energies $E$. Then a random walk is performed in energy space by flipping spins randomly with a transition probability of
\begin{align}
p(E_1\rightarrow E_2) = min\left(\frac{g(E_1)}{g(E_2)},1\right),
\end{align}
where $E_1$ and $E_2$ are energies before and after a spin is flipped, while simultaneously augmenting the density of states $g(E)\rightarrow g(E)*f$ by a multiplicative factor $f>1$ and incrementing a histogram $H(E)$ of visited configurations. By construction, more probable (higher entropy) energy levels develop higher $g(E)$, and transition probabilities level out, producing a flat histogram. The factor $f$ starts with a high value, such as $f_0=2.718$ ($\sim e$, the base of natural logarithms), and is reduced according to some schedule whenever the histogram fulfills a flatness criterion (e.g., $H(E)$ is not less than $80\%$ of $\langle H(E)\rangle$ for all possible $E$), after which the histogram is reset $H(E)=0$. The simulation process is repeated until a lower bound of $f$ is reached (typically, $f_{min} = 10^{-8}$). The appendix details parameter choices made for this work.

\section{Simulations}
Using the above defined formulation, we map a range of deep neural networks onto Ising models and analyze their spin possible configurations, that is, their density of states. This work focuses on transformers~\cite{vaswani2017}, as they have spurred much interest in recent years, and grown to billions of weights, consuming vasts amounts of resources to train. Pretrained transformers are retrieved from Huggingface (\url{https://huggingface.co/}), covering both encoder- and decoder-based transformers of various sizes ($L\in[12,24]$ and $H\in[768,1024,2048]$) for language and vision tasks, including: GPT2~\cite{gpt2}, OPT~\cite{opt}, Bloom~\cite{bloom}, BERT~\cite{bert}, BEiT~\cite{beit}, DeiT~\cite{deit}, ViT~\cite{vit}. We construct the Ising models and compute their density of states using weight from trained networks, and draw comparisons to networks that take ``random" weights in order to quantify what properties might appear after training. These pseudo-random models are built by shuffling the trained weights, rather than taking their values at initialization, as weights distributions can change after training, thus impacting the energy magnitudes. More concretely, shuffling swaps each network weight $w^\ell_{ij}$ with a randomly chosen weight $w^{\ell^\prime}_{i^\prime j^\prime}$, such that all weights are swapped at least once, and repeating this entire process ten times.

\begin{table}[!b]
\caption{Minimum energy $E_{min}$ and width $W$ of the density of states $g(E)$ for various trained neural networks and after shuffling all $B$ bonds, normalized by the number of spins $N$.}
\centering
\resizebox{\columnwidth}{!}{
\begin{tabular}{lrrrrrrrrrrrr}
\hline
\hline
        &&      &&          &&  \multicolumn{3}{c}{Trained} && \multicolumn{3}{c}{Shuffled} \\
Network &&  $B (10^6)$ &&    $N$   &&  $E_{min}/N$ &&  $W/N$       && $E_{min}/N$ &&  $W/N$        \\
\hline
%model      bonds
%opt-125m   70778880    71 * 1e6
%opt-350m   251658240   252 * 1e6
%opt-1.3b   1006632960  1007 * 1e6
vit-base    &&  71   &&  65,280     &&  -2.33   && 8.84 &&  -1.07   && 1.89 \\
deit-base   &&  71   &&  65,280     &&  -0.24	&& 0.44 &&  -0.19   &&  0.43 \\
beit-base   &&  71   &&  65,280     &&  -0.43	&& 1.14 &&  -0.38   &&  0.86 \\
bert-base   &&  71   &&  65,280     &&  -2.01	&& 2.74 &&  -0.52   &&  0.90  \\
bert-large  &&  252  &&  173,056    &&  -0.25	&& 0.42 &&  -0.16   &&  0.25  \\
opt-125m    &&  71   &&  65,280     &&  -2.35	&& 3.02 &&  -0.56   && 0.97  \\
opt-350m    &&  252  &&  173,056    &&  -0.47	&& 0.88 &&  -0.08   && 0.22  \\
opt-1.3b    &&  1007 &&  346,112    &&  -0.23	&& 0.27 &&  -0.04   &&  0.09   \\
bloom-350m  &&  252  &&  173,056    &&  -0.10   && 0.22 &&  -0.10   && 0.19 \\
gpt2        &&  71   &&  65,280     &&  -1.79   && 6.96 &&  -1.67   && 3.0 \\
gpt2-medium &&  252  &&  173,056    &&  -0.33   && 0.92 &&  -0.52   && 0.82 \\

\hline
\hline
\end{tabular}
}
\label{tabdos}
\end{table}

Figure~\ref{figdos} plots $g(E)$ for various transformers, where the energy curves differ substantially between the trained and shuffled weights. More specifically, networks that have been trained observe a wider $g(E)$, where the width is given by $W=E_{max}-E_{min}$, which means there is a large dispersion in energies between realizable configurations. On the other hand, shuffling the weights substantially diminishes $W$, or the range of energies that different spin configurations can achieve. For instance, $g(E)$ for bert-base spans $2.74$ energy values per spin after training compared to $0.9$ after shuffling. Ising models constructed from trained weights also achieve substantially lower ground states $E_{min}$ than from shuffled weights. For example, opt-1.3b achieves an energy minimum of $-0.23$ per spin compared to $-0.04$ after shuffling. Similar conclusions can be made for other transformer models, as shown in Table~\ref{tabdos} and corresponding $g(E)$ curves in the appendix.

These results show that networks behave widely different after training compared to taking a random ordering of their values, which can only arise from structure, or how weights are arranged in a neural network. One plausible explanation is that they learn specific structures throughout training that makes them more amenable for achieving low-energy states. More specifically, combinations of large weights will determine the value of $E_{min}$, as they contribute the most to the energy (e.g., an energy difference of $\Delta E=w_1w_2$ arises from a weight pair $w_1,w_2$ connected with a neuron with additive contributions), whereas small weights have a negligible impact. Furthermore, the emergence of structures across neural networks of various sizes and trained on widely different data for distinct tasks suggests they represent a ``universal" property to learning.

\begin{figure}[t]
\begin{tabular}{cc}
\includegraphics[width=\linewidth]{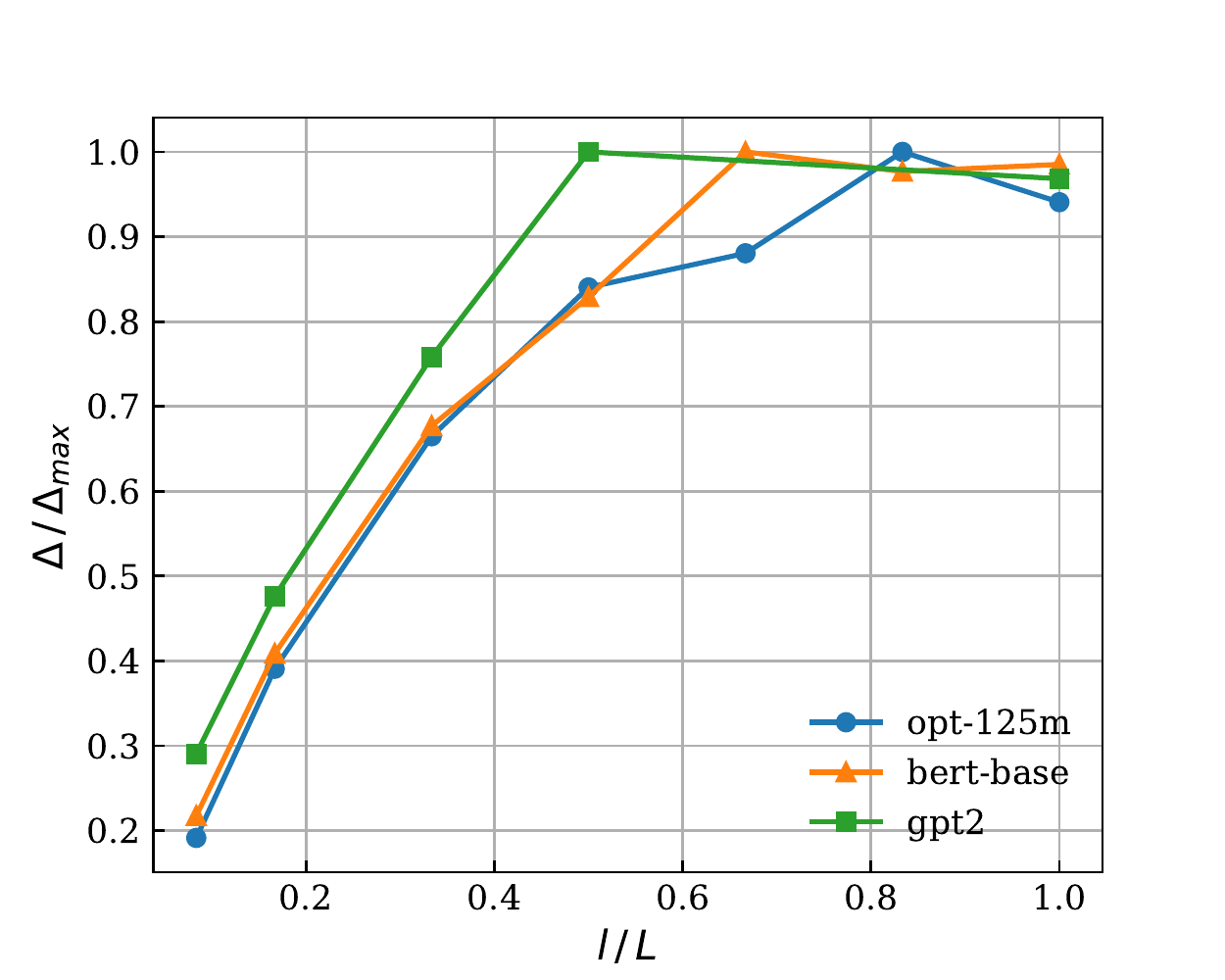}
\end{tabular}
\caption{Difference in $g(E)$ widths for trained and shuffled, $\Delta=W_\text{train}-W_\text{shuffled}$, as a function of the number of transformer layers $l$ that participate in the Ising model, normalized by the maximum $L$. Colors denote the various neural networks considered: opt-125m, bert-base, and gpt2.}
\label{figlayers}
\end{figure}

To determine whether the observed structures appear across the entire network or are specific to a few layers, Ising models are constructed using a subset of transformer layers (each consisting of four weight matrices). Figure~\ref{figlayers} shows the difference in $g(E)$ widths between trained and shuffled, $\Delta=W_{\text{train}}-W_{\text{shuffle}}$, for different amounts of transformer layers $l$. It can be seen that $\Delta$ increases as a function of layers for all networks considered. While the trained $g(E)$ approaches that of shuffled which has no structure (i.e., $\Delta=0$) when using few layers, adding more layers shows a clear distinctions between weights that have been trained and shuffled. More specifically, $\Delta$ saturates once half of the network layers are included ($l/L=0.5$), which suggests that structures span most of the network after training and cannot be observed in a single transformer layer, or even worse a single weight matrix, as evaluated in previous works.

\begin{figure}[!t]
\begin{tabular}{cc}
\includegraphics[width=\linewidth]{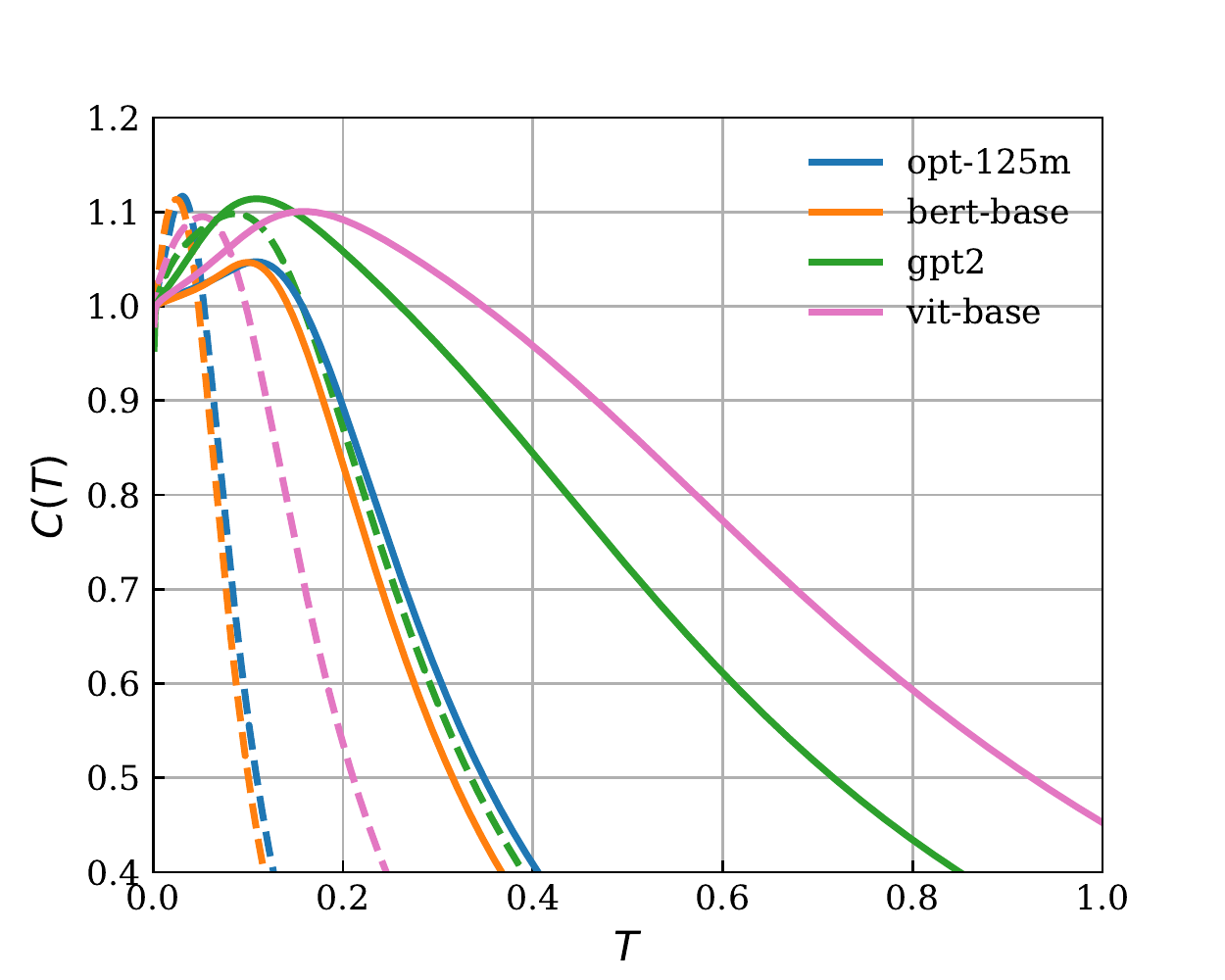}
\end{tabular}
\caption{Specific heat $C(T)$ as a function of temperature $T$ (in arbitrary units) for various neural networks: gpt2, opt-125m, bert-base, and vit-base, denoted by different colors. Solid and dotted lines represent trained and shuffled, respectively.}
\label{figspecificheat}
\end{figure}

\begin{table}[!b]
\caption{Density of states width $W/N$ and task error $\mathcal{E}$ (tasks are denoted in subscript) across various networks using different fractions $f$ of values being shuffled.}
\centering
\resizebox{\columnwidth}{!}{\begin{tabular}{llrrrrrrrrrrrrrrr}
\hline
\hline
          && $f$ && $0$ && $0.01$ && $0.05$ && $0.1$ && $0.2$ && $0.5$ && $1$ \\
\hline
bert-base && $W/N$ && 2.73 && 2.38 && 1.69 && 1.20 && 0.86 && 0.85 && 0.90 \\
\hline
opt-125m  && $W/N$ && 2.74 && 2.29 && 1.59 && 1.12 && 0.92 && 0.93 && 0.92 \\
          && $\mathcal{E}_{\text{Wikitext2}}$ && $0$ && $0.02$ && $1.5$ && $4.4$ && $31.5$ && $11435$ && - \\
          && $\mathcal{E}_{\text{Wikitext103}}$ && $0$ && $0.03$ && $1.5$ && $6.0$ && $29.9$ && $14893$ && - \\
          && $\mathcal{E}_{\text{Lambada}}$ && $0$ && $0.03$ && $1.3$ && $5.2$ && $41.4$ && $4817$ && - \\
\hline
gpt2      && $W/N$ && 6.96 && 6.07 && 4.50 && 3.40 && 3.00 && 2.96 && 3.00 \\
          && $\mathcal{E}_{\text{Wikitext2}}$ && $0$ && $0.03$ && $0.3$ && $3.1$ && $30.7$ && $9808$ && - \\
          && $\mathcal{E}_{\text{Wikitext103}}$ && $0$ && $0.02$ && $0.6$ && $4.2$ && $29.4$ && $10059$ && - \\
          && $\mathcal{E}_{\text{Lambada}}$ && $0$ && $0.03$ && $1.3$ && $5.2$ && $41.4$ && $4817$ && - \\
\hline
vit-base  && $W/N$ && 8.84 && 7.88 && 5.73 && 4.01 && 2.04 && 1.90 && 1.89 \\
          && $\mathcal{E}_{\text{Imagenet}}$ && $0$ && $0.02$ && $0.01$ && $0.1$ && $0.5$ && $15$ && $79$ \\
\hline
\hline
\end{tabular}}
\label{tabacc}
\end{table}

So far, trained networks have been compared to their values after shuffling, where the latter performs equivalently to a random network. Thus, an interesting question is whether the density of states distinguishes between networks that achieve different performance (e.g., from different stages of training), which can be approximated by varying the fraction $f$ of values that are shuffled. Table~\ref{tabacc} shows the width in the density of states for different amounts of shuffling, where we observe that $W/N$ approximates that of a random network ($f=1$) when more than $20\%$ of values have been shuffled.

To correlate energy values to performance, transformers are evaluated on language or vision tasks after shuffling their weights. For language tasks, text generation is considered on Wikitext-2, Wikitext-103~\cite{wikitext}, and Lambada~\cite{lambada}, where perplexity evaluates network quality. For vision tasks, networks are evaluated for image classification on Imagenet~\cite{imagenet} comprising $1k$ possible classes, where  classification accuracy is computed using $10k$ images sampled from the validation set. The task errors $\mathcal{E}$ are computed as the relative change in evaluation metrics, $\mathcal{M}$, between trained and shuffled networks, $\mathcal{E}=\frac{|\mathcal{M}_{\text{trained}}-\mathcal{M}_{\text{shuffled}}|}{\mathcal{M}_{\text{trained}}}\times100$. Table~\ref{tabacc} shows that $\mathcal{E}$ increases with shuffling; errors above $1$ are considered substantial in the deep learning community. Interestingly, the energy values, or more specifically the widths $W$, are much more sensitive to changes in weights than the task errors $\mathcal{E}$. From a more practical perspective, this implies that the density of states could be used to evaluate how well have neural networks been trained \textbf{\textit {without using any external data}}. This is increasingly important as networks become multi-purpose, where obtaining representative evaluation data for every possible task is challenging (e.g., in language domains, hundreds of tasks are often needed to determine network quality~\cite{brown2020}).

From the density of states, one can derive various thermodynamic quantities, such as specific heat. The specific heat can be defined through the expression:
\begin{align}
C(T) = \frac{\langle E^2\rangle_T - \langle E\rangle_T^2}{T^2},
\end{align}
where $\langle f\rangle=\sum_E f g(E)e^{-\beta E}/\sum_Eg(E)e^{-\beta E}$ denotes the expectation given $\beta=1/k_BT$. Figure~\ref{figspecificheat} shows that $C(T)$ differs across networks, but some such as bert-base and opt-125m exhibit similar curves. After shuffling, $C(T)$ shifts to the left, dropping to zero at much lower temperatures $T$, and achieving lower critical temperature $T_C$ (i.e., temperature that maximizes $C(T)$). For example, opt-125m has $T_c=0.11$, which is $3\times$ higher than $T_c=0.03$ obtained after shuffling its weights. Similar conclusions can be made for the other networks, as detailed in the appendix.
These phenomenological findings require more studies for better understanding of their implications.

\section{Conclusion}
The current work analyzes deep neural networks from a statistical physics (thermodynamics) perspective, where weights are mapped to exchange interactions, and nodes to spin glass Ising model spins. By calculating the density of states, we demonstrate that structures emerge in weight values after training. For example, well-trained networks span a much wider range of energies than can be realized on poorly trained networks. This work opens several avenues for future research. One direction should focus on analyzing other thermodynamic quantities which may provide further insights into what properties neural networks obtain after training. For this purpose, the  density of states have been released at \url{https://github.com/stosicresearch/dnnising}. Another direction is to expand on its practical applications, such as: distinguishing between weights during training, thereby serving as a metric when to pause the training procedure; determining network quality for various parameter choices; comparing networks that were trained on different data for distinct tasks; to name a few.

%\bigskip\noindent{\bf Acknowledgments}\\

\bibliography{dnnising} 

\newpage

\section{Simulation Details}\label{appendix:wl}
This appendix details parameter choices made for the Wang-Landau simulations. The simulation executes a total of $10N$ spin flips before normalizing and resetting the histogram $H(E)$, where the number of updates increases by $10\%$ after every reset. The multiplicative factor $f$ follows a piecewise schedule with different slopes on a logarithmic scale to promote quicker convergence of the density of states early in the simulation, allowing for finer tuning of $g(E)$ in the late stages. Lastly, the simulation is paused when $f$ falls below $10^{-4}$ since convergence did not improve for smaller thresholds.

The limitations of the current work are also worth mentioning. As the problem size increases, the density of states becomes notoriously difficult to obtain due to the explosion in the number of possible configurations, since the probability for a random walk to visit configurations of lower energy (and thus lower occurrence) goes down drastically. Ref.~\cite{wanglandau} has been able to compute accurate density of states up to a square lattice of $256\times256$ spins. For comparison, the neural networks considered span between  65,280 and 346,112 spins which would map to $255\times255$ and $588\times588$ on a square lattice, matching the problem size previously studied. However, since these networks make up at most $1.3$ billion weights (roughly the number of bonds), this also means that state-of-the-art networks used in industry, spanning hundreds of billions of weights, would represent over a $100$-fold increase in problem size. As a result, the current techniques cannot be used to study neural networks at the largest scales, since their large number of spins would make it difficult to achieve good convergence. Nevertheless, from a practical angle, the density of states estimate obtained through simulation could still provide meaningful differences between networks of varying quality, even if it's not fully converged.

\section{Density of States}\label{appendix:dos}
\begin{figure*}[t]
\begin{tabular}{c}
\includegraphics[width=\linewidth]{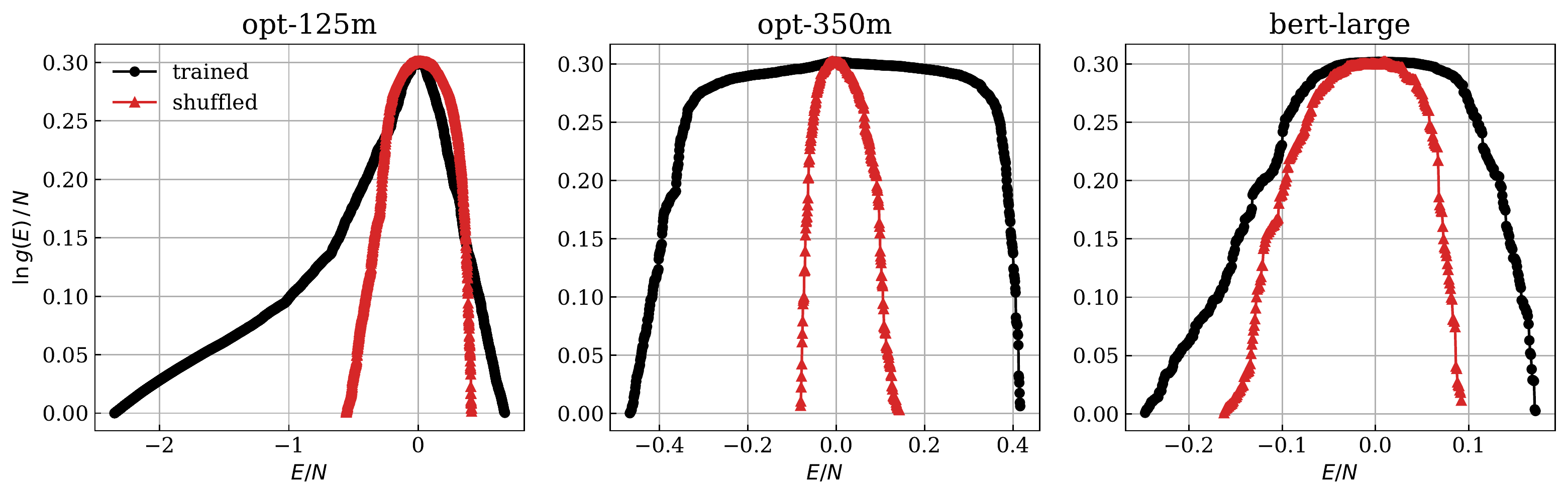}
\end{tabular}
\caption{The density of states, $g(E)$, for various language transformers before and after shuffling.}
\label{figdosappendix1}
\end{figure*}

\begin{figure*}[htb]
\begin{tabular}{c}
\includegraphics[width=\linewidth]{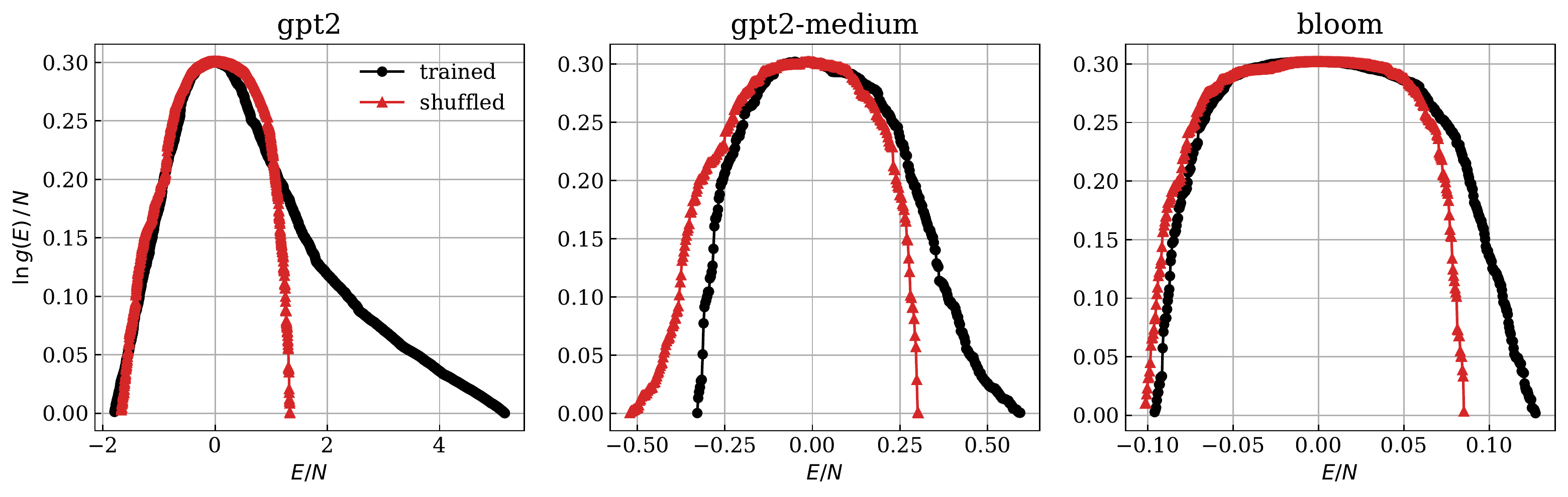}
\end{tabular}
\caption{The density of states, $g(E)$, for various language transformers before and after shuffling.}
\label{figdosappendix2}
\end{figure*}

\begin{figure*}[htb]
\begin{tabular}{c}
\includegraphics[width=\linewidth]{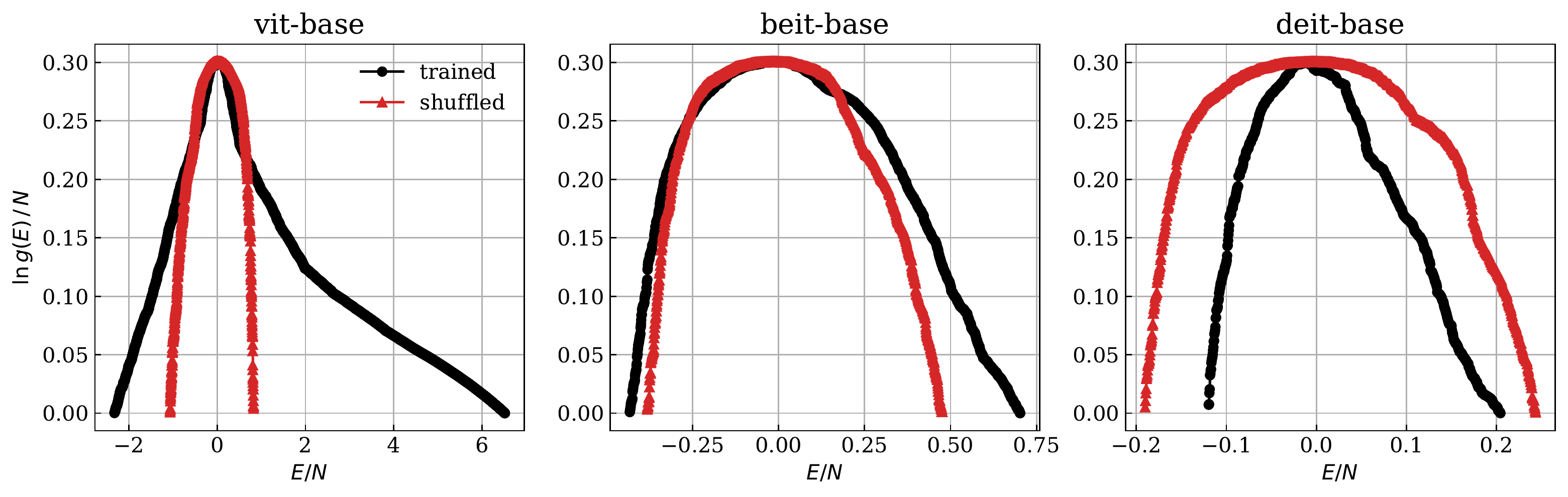}
\end{tabular}
\caption{The density of states, $g(E)$, for various vision transformers before and after shuffling.}
\label{figdosappendix3}
\end{figure*}

This appendix extends the density of states for the remaining neural networks studied. Figures~\ref{figdosappendix1}-\ref{figdosappendix3} illustrate $g(E)$ for a suite of pretrained transformers used for language and vision. In most cases, the density of states covers a much wider range of energies for trained than shuffled, with a few situations where that difference is less obvious. Noteworthy exceptions include deit-base, where the trained network achieves a narrower $g(E)$ than after shuffling, which we suspect might be due to poor convergence of the Wang-Landau simulation, as described in the previous appendix section.

\section{Specific Heat}\label{appendix:tc}
This appendix expands on the specific heat computations. To deal with numerical stability, the energy contributions for $\langle E^2\rangle$ and $\langle E\rangle^2$ are calculated using
\begin{align}
p(E_i)=e^{-(E_i/2B - E_{min}) / T + \ln g(E_i)/N - g_{max}},
\end{align}
where $E_{min}=\max(E/2B)$ and $g_{max}=\max(\ln g(E)/N)$ avoids overflow from having big values in the exponential. By normalizing $E$ with the number of bonds $B$ and $g(E)$ with the number of spins $N$, the temperature is in units of $k_b/J=1$, which has been confirmed to produce correct $C(T)$ for an $8\times8$ Ising model, where the maximum occurs at $k_bT/J=2.269$ (where $k_b$ is the Boltzmann constant) as expected.

\begin{figure*}[t]
\begin{tabular}{c}
\includegraphics[width=\linewidth]{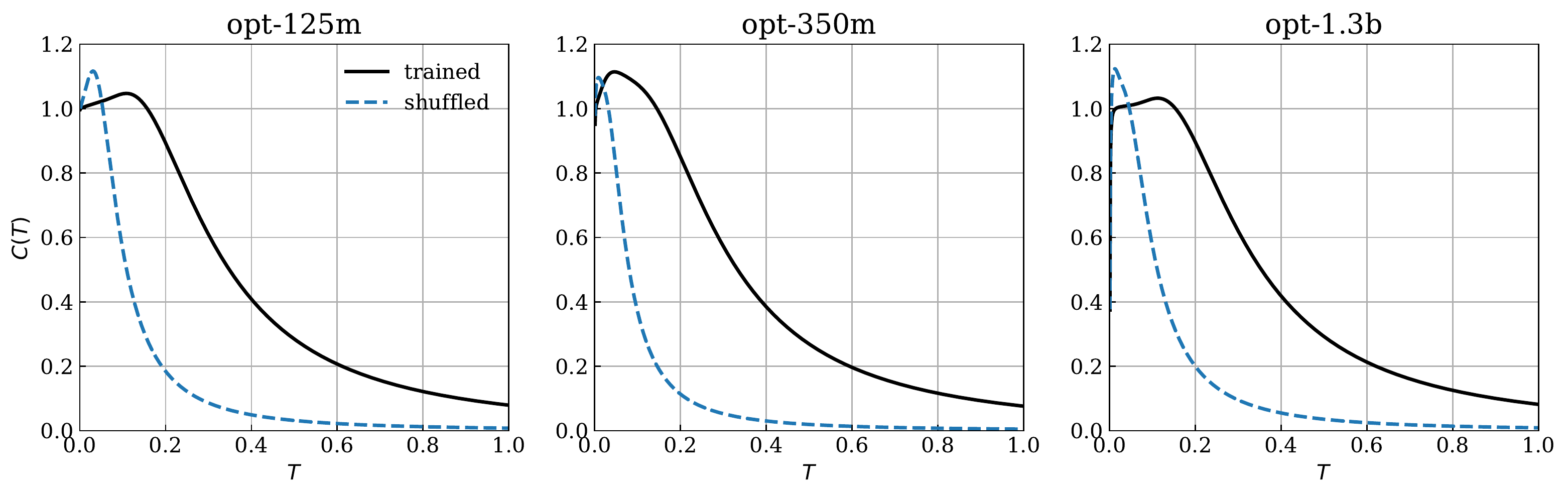}
\end{tabular}
\caption{The specific heat, $C(T)$, for various language transformers before and after shuffling.}
\label{figspecificheatappendix1}
\end{figure*}

\begin{figure*}[t]
\begin{tabular}{c}
\includegraphics[width=\linewidth]{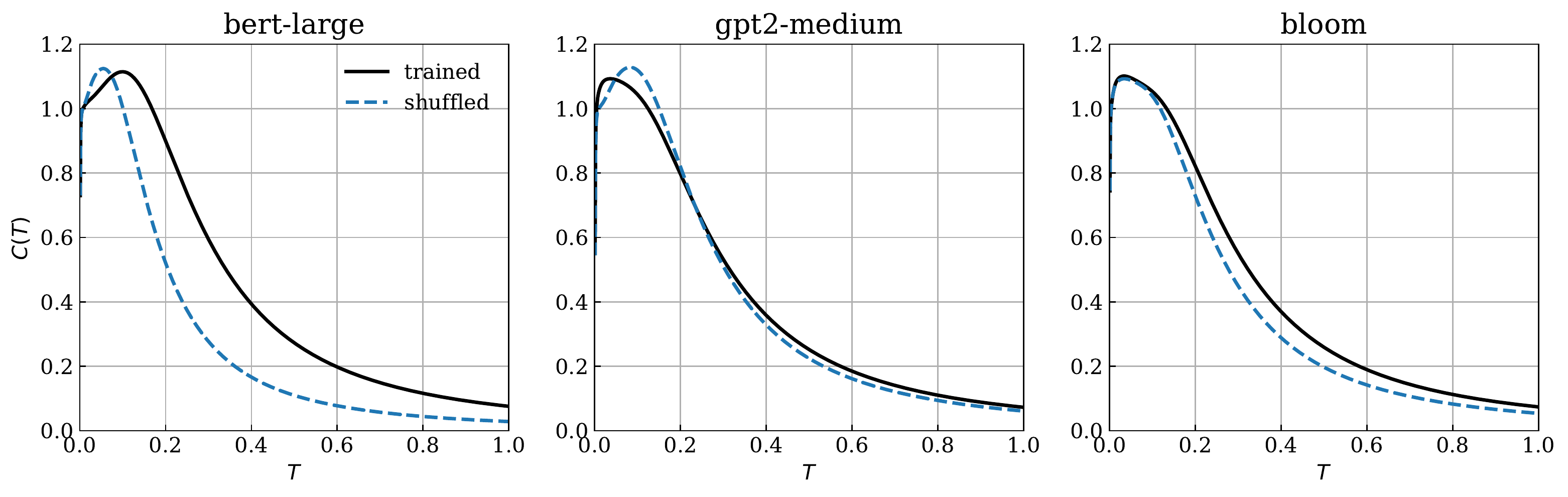}
\end{tabular}
\caption{The specific heat, $C(T)$, for various language transformers before and after shuffling.}
\label{figspecificheatappendix2}
\end{figure*}

\begin{figure*}[t]
\begin{tabular}{c}
\includegraphics[width=\linewidth]{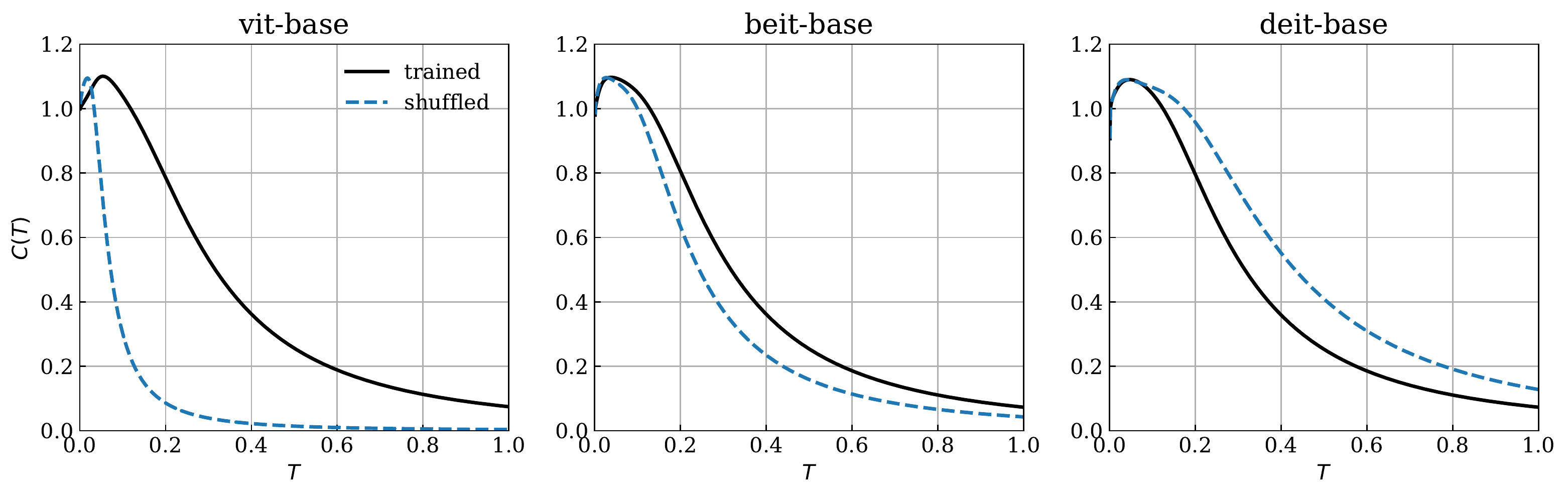}
\end{tabular}
\caption{The specific heat, $C(T)$, for various vision transformers before and after shuffling.}
\label{figspecificheatappendix3}
\end{figure*}

While the number of bonds is proportional to the number of spins for an Ising model on a square lattice (4n for 2d and  6n for 3d nearest neighbors), for neural networks that number grows quadratically. Combined with the fact that the weight values (or bonds) are unbounded, the range of energies grows immensly (e.g., the maximum for an $8\times8$ Ising model is $128J$ compared to $150,000J$ for a neural network). As a result, $p(E_i)$ values cannot all be represented using floating point precision (e.g., $p(E_i)$ covers values from $e^{-454074}$ to $e^{+74}$ at $T=1$, which is well beyond the $2^{2048}$ range of values in double precision), even after exploring countless ``numerical tricks" and extended precision formats. The solution adopted in this work was to normalize the energies in the $[0,1]$ range using $(E-E_{min})/(E_{max}-E_{min})$, where the scale for $T$ is in ``arbitrary units". However, care must be taken in the normalization when comparing $C(t)$ across different networks. More specifically, Figure~\ref{figspecificheat} in the paper uses $E_{min}$ and $E_{max}$ from gpt-125m to normalize energy values for the remaining networks, which guarantees they all have the same scales for $T$. On the other hand, Figures~\ref{figspecificheatappendix1}-\ref{figspecificheatappendix3} normalizes energies for trained and shuffled using $E_{min}$ and $E_{max}$ from each network individually.

Using the specific heat, the critical temperature $T_c$ is obtained from the value of $T$ that maximizes $C(T)$. Table~\ref{tabspecificheat} lists $T_c$ values achieved for various trained and shuffled networks. It can be observed that $T_c$ changes after shuffling, but it can't be compared between networks because the scales vary. 
\begin{table}[!htb]
\caption{Critical temperature $T_c$ obtained for trained neural networks and after their values have been shuffled.}
\centering
\resizebox{\columnwidth}{!}{
\begin{tabular}{lrrrrrrrrrr}
\hline
\hline
Network			&&	Trained $T_c$	&& Shuffled $T_c$	&&	Network	&&	Trained $T_c$	&& Shuffled $T_c$	\\
\hline
opt-125m 		&&	0.109	&&	0.031	&&	bert-base	&&  0.11	&&	0.027	\\
opt-350m 		&&	0.046	&&	0.009	&&	bert-large	&&  0.1	    &&	0.055	\\
opt-1.3b 		&&	0.112	&&	0.013	&&	vit-base	&&  0.054	&&	0.018	\\
gpt2.dat		&&	0.047	&&	0.038	&&	deit-base	&&  0.048	&&	0.04	\\
gpt2-medium	    &&	0.036	&&	0.083	&&	beit-base	&&  0.039	&&	0.027	\\
bloom			&&	0.034	&&	0.034	\\
\hline
\hline
\end{tabular}
}
\label{tabspecificheat}
%\end{ruledtabular}
\end{table}

\end{document}